\documentclass[12pt,preprint]{aastex}
\slugcomment{Accepted by ApJ. Letters, March 17, 2004}
\usepackage{natbib}
\begin{document}
\title{Imaging the Disk around TW Hya with the Submillimeter Array}
\author{Chunhua~Qi\altaffilmark{1}, Paul~T.P.~Ho\altaffilmark{1},
David~J.~Wilner\altaffilmark{1}, Shigehisa~Takakuwa\altaffilmark{2},
Naomi~Hirano\altaffilmark{3}, Nagayoshi~Ohashi\altaffilmark{2,3},
Tyler~L.~Bourke\altaffilmark{2}, Qizhou~Zhang\altaffilmark{1},
Geoffrey~A.~Blake\altaffilmark{4}, Michiel~Hogerheijde\altaffilmark{5},
Masao~Saito\altaffilmark{2,6}, Minho~Choi\altaffilmark{3,8}, \&
Ji~Yang\altaffilmark{7}}

\altaffiltext{1}{Harvard--Smithsonian Center for
Astrophysics, 60 Garden Street, MS 42, Cambridge, MA 02138,
USA; cqi@cfa.harvard.edu, ho@cfa.harvard.edu, dwilner@cfa.harvard.edu,
qzhang@cfa.harvard.edu.} 
\altaffiltext{2}{Harvard--Smithsonian Center for
Astrophysics, Submillimeter Array Project, 645 N. A'ohoku Place, Hilo,
HI 96720, USA; stakakuw@sma.hawaii.edu, tbourke@sma.hawaii.edu, nohashi@sma.hawaii.edu.}
\altaffiltext{3}{Academia Sinica Institute of Astronomy \& 
Astrophysics, P.O. Box 23-141, Taipei, Taiwan, 106, R.O.C.; hirano@asiaa.sinica.edu.tw.}
\altaffiltext{4}{Divisions of Geological \& Planetary Sciences and
Chemistry \& Chemical Engineering, California Institute of Technology,
MS150--21, Pasadena, CA 91125, USA; gab@gsp.caltech.edu.}
\altaffiltext{5}{Sterrewacht Leiden, P.O. Box 9513, 
2300 RA Leiden, The Netherlands; michiel@strw.leidenuniv.nl.}
\altaffiltext{6}{National Astronomical Observatory of Japan,
2-21-1 Osawa, Mitaka, Tokyo 181-8588, Japan; Masao.Saito@nao.ac.jp.}
\altaffiltext{7}{Purple Mountain Observatory, Chinese Academy of Sciences,
Nanjing 21008, China; jiyang@pmo.ac.cn.}
\altaffiltext{8}{Taeduk Radio Astronomy Observatory, 
Korea Astronomy Observatory,Hwaam 61-1, Yuseong, Daejeon 305-348,
Korea; minho@cosmos.trao.re.kr.}

\begin{abstract}
We present $\sim$2$''$-4$''$ aperture synthesis observations of the
circumstellar disk surrounding the nearby young star TW Hya in the CO
J=2--1 and J=3--2 lines and associated dust continuum obtained with
the partially completed Submillimeter Array.  The extent and peak
flux of the 230 and 345 GHz dust emission follow closely the 
predictions of the irradiated accretion disk model of 
\citet{calvet_d02}. 
The resolved molecular line emission extends to a radius of at
least 200 AU, the full extent of the disk visible in scattered light,
and shows a clear pattern of Keplerian rotation. Comparison of
the images with 2D Monte Carlo models
constrains the disk inclination angle to $7^{\circ}\pm 1^{\circ}$. 
The CO emission is optically thick in both lines, and the kinetic
temperature in the line formation region is $\sim$20~K. Substantial
CO depletion, by an order of magnitude or more 
from canonical dark cloud values, is required to
explain the characteristics of the line emission.

\end{abstract}


\section{Introduction}

TW Hya is the closest known classical T Tauri star, and exhibits
high X-ray flux, large lithium abundance, and evidence for an 
actively accreting disk with accretion rate estimated to be 
from 10$^{-9}$ to 10$^{-8}$ M$_{\odot}$ yr$^{-1}$ based on the
large H${\alpha}$ equivalent width and the X-ray emission 
(\citealp{kastner_h02} and refs therein).
Despite its apparently advanced age, estimated to be 5 to 20 Myr, TW Hya 
is surrounded by a disk of mass $\sim 5 \times 10^{-3}$ M$_{\odot}$ 
inferred from dust emission (\citealp{wilner_h00}). The disk is
viewed nearly face-on, and is visible in scattered light to a
radius of at least $3\farcs5$, or 200 AU at a distance of 56 pc 
 (\citealp{krist_s00, trilling_k01, weinberger_b02, calvet_d02}).

Several molecular species (CO, HCN, CN, HCO$^+$, DCO$^+$) have
been detected in the TW Hya disk at millimeter and submillimeter wavelengths 
using single dish telescopes (\citealp{kastner_z97, vanzadelhoff_v01, thi01, 
vandishoeck_t03}). The proximity, isolation, and rich chemistry of
the TW Hya disk make this system an excellent target for the study of
the physical and chemical structure of protoplanetary environments 
at high spatial resolution using interferometry.  Models may be directly 
compared to aperture synthesis images of molecular lines to verify the 
inferences from low resolution observations, 
and to further probe the kinematics, 
physical conditions, and chemistry. A first step in this direction was the
imaging analysis of HCO$^+$ J=1--0 emission observed with the Australia 
Telescope Compact Array by ~\citet{wilner_b03}, though the spatial and 
spectral resolution were insufficient to examine the kinematics in any detail.

The prospects are excellent for interferometric observations of protoplanetary
disks at submillimeter wavelengths (\citealp{blake02}). The Submillimeter Array
(SMA)\footnote{The Submillimeter Array is a joint
project between the Smithsonian Astrophysical Observatory and the
Academia Sinica Institute of Astronomy and Astrophysics, and is
funded by the Smithsonian Institution and the Academia Sinica.}
, now nearing completion on Mauna Kea, is an ideal instrument for making 
these observations (\citealp{ho_m04}). 
In this {\em Letter}, we present early SMA observations of
the TW Hya disk in the CO J=3--2 and J=2--1 lines and dust continuum, with a
maximum resolution of $2''$. The resulting images unambiguously reveal the
rotation of the disk around the star and constrain the disk inclination
and size.
 
\section{Observations}

The SMA observations of TW Hya were made between 2003 March and June 
using two configurations of five of the 6 meter diameter antennas 
at each of two frequencies (230 GHz and 345 GHz). These observations 
provided 17 independent baselines ranging in length from 16 to 120 meters. 
Table~\ref{tab:obs} summarizes the observational parameters. The
synthesized beam sizes were $\sim$2$''$ at 345 GHz and $\sim$4$''$ at 230 GHz
with robust weighting.  The SMA digital correlator was configured
with both a narrow band of 512 channels over 104 MHz, which provided
0.2 MHz frequency resolution, or 0.18 km~s$^{-1}$ 
velocity resolution at 345 GHz, and several broader bands that provided 
656 MHz of bandwidth for continuum measurements.  Calibration of the 
visibility phases and amplitudes was achieved with observations of the 
quasar 3C273, typically at intervals of 30 minutes. Two weaker quasars
closer to TW Hya on the sky, 1058+015 and 1037-295, were also observed to
verify the effectiveness of phase referencing between 3C273 and TW Hya. 
Observations of Callisto provided the absolute scale for the flux density
calibration. The uncertainties in the flux scale are estimated to be
20\% based on the scatter of the measurements on different days, which is
due to the uncertainties in pointing models and system temperature 
measurements. 
The data were calibrated using the MIR software package adapted for the SMA
from the IDL version of the MMA software package developed originally for 
OVRO (Scoville et al. 1993).  The continuum emission from TW Hya is strong 
enough for self-calibration, and an iteration of phase-only self-calibration
was performed to improve the images by 20\%. 
The NRAO AIPS package was
used to generate continuum and spectral line images, all of which were
CLEANed in the usual way. 

\section{Results}

\subsection{Continuum Emission}

\begin{figure}
\centering
\includegraphics[width=3in, angle=-90]{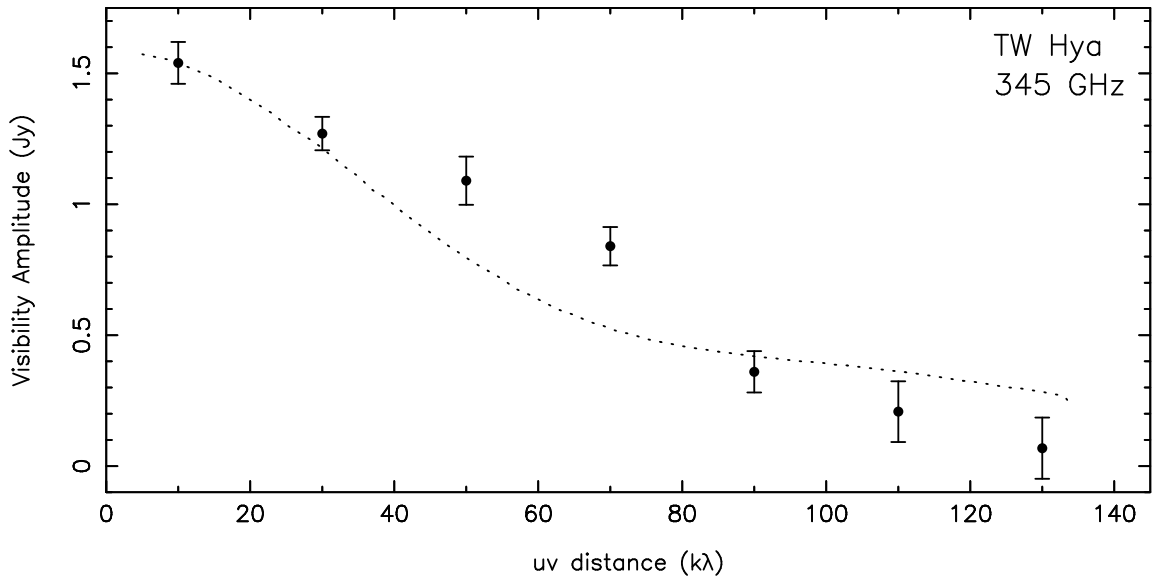}
\caption{ Visibility amplitudes of the observed and modeled 345 GHz 
continuum emission from TW Hya, annularly averaged in 20~k$\lambda$ bins.
The error bars represent $\pm1$ standard deviation for each bin.
The curve shows the visibility amplitudes derived from the model
calculations of Calvet et al. (2002) of an irradiated accretion
disk, which fits the full spectral energy distribution well after
applying an amplitude scale factor of 0.9.
\label{fig:profile}}
\end{figure}

\citet{calvet_d02} have developed a physically self-consistent
irradiated accretion disk model for TW Hya. This unique model requires a 
distinct transition in disk properties at a radius of $\approx$4~AU, 
likely due to a developing gap, and grains that have grown to sizes 
of $\approx$1--10 cm in order to match the full spectral energy 
distribution from optical to radio wavelengths.
The Calvet model is consistent with the radial brightness 
distribution of the disk observed at 43~GHz, where the dust emission 
has been resolved by the VLA (\citealp{wilner_h00}). 
The new SMA continuum measurements agree well with previous single dish 
observations at these wavelengths, and the predictions of the
Calvet model.
To compare the dust emission to the model radial brightness, 
we show in Figure~\ref{fig:profile} the visibility amplitudes at 345 GHz, 
annularly averaged in 20~k$\lambda$ bins, with the model curve
(scaled by 0.9, well within the uncertainties of the flux scale).
The good match of the SMA continuum measurements with previous data and 
the irradiated accretion disk model predictions gives confidence in the
reliability of the submillimeter measurements, as well as the accuracy
of the absolute flux scale.

\subsection{CO J=2--1 and J=3--2 Emission}

\begin{figure}
\epsscale{1.0} \plotone{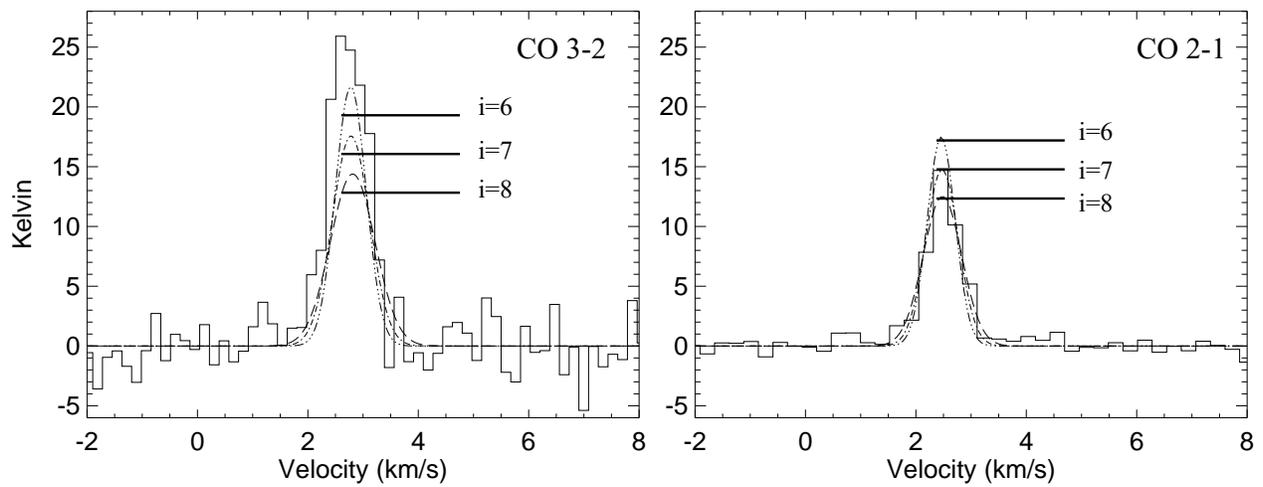}
\caption{
The CO J=3--2 and J=2--1 spectra at the continuum (stellar) position, 
beam sizes are listed in Table~1. The curves show synthetic spectra 
for a small range of inclination angles using the model parameters 
listed in Table 2.
 \label{fig:spectra}}
\end{figure}

Figure~\ref{fig:spectra} shows spectra of the CO J=2--1 and J=3--2
line emission at the TW Hya continuum peak. 
Both lines are detected with high signal-to-noise ratios.
The CO emission from the disk is spatially resolved; a circular 
Gaussian fit to the CO 3--2 integrated intensity map
gives a FWHM size of $3\farcs5\pm0\farcs3$. 
Table~\ref{tab:obs} lists the integrated intensities and results 
of Gaussian fits to the central spectra. 
As indicated by \citet{beckwith_s93} and \citet{vanzadelhoff_v01},
CO J=2--1 and CO J=3--2 emission are expected to be highly optically
thick in circumstellar disks. These lines are
thus excellent probes of the outer disk velocity field but do not
robustly trace the disk mass (\citealp{simon_d01, qi_k03}).
Further, the peak intensities should be tied to the kinetic 
temperatures of the regions of the disk where
the lines become optically thick, and as TW Hya appears almost face
on, any differences in brightness may be related to vertical
temperature gradients.

\begin{figure}
\centering
\includegraphics[width=5in,angle=-90]{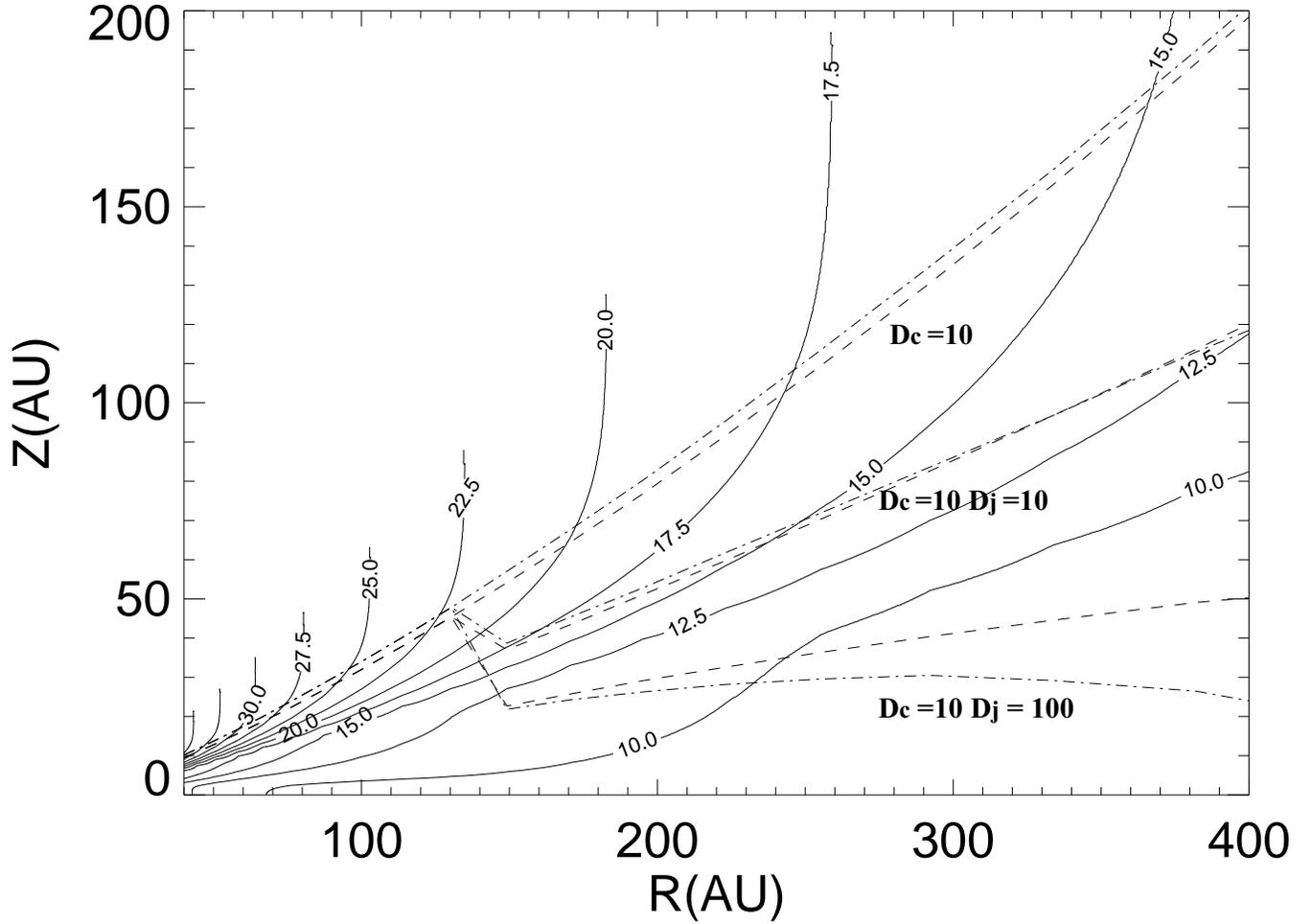}
\caption{
The solid contours show the gas kinetic temperature from the TW Hya model
of Calvet et al. (2002). The dashed and dot-dashed lines show the  
$\tau=1$ surfaces for the CO J=3--2 and J=2--1 lines for the three 
depletion scenarios described in the text.
 \label{fig:tau}}
\end{figure}

Comparing the intensities of the two transitions within the same region 
(here the CO J=3--2 data set is convolved with the CO J=2--1 beam), 
the peak brightness of the CO J=3--2 line is $21.0\pm0.7$ K 
and that of the CO J=2--1 line is $14.7\pm0.4$ K. 
Figure~\ref{fig:tau} shows the location of the $\tau=1$ surfaces for 
these CO transitions calculated together with contours from the kinetic 
temperature model of Calvet et al. (2002) for three
different depletion scenarios. All assume a constant depletion factor
(D$_c$) of 10. The top set is calculated with the constant D$_c$,
the one below includes additional jump depletion (D$_j$) by a factor of 10 
where the kinetic temperature falls below 22~K (the critical temperature of
CO desorption, \citealp{vanzadelhoff_v01}),
and the bottom one includes D$_j$ of 100 at
that same temperature threshold. For the first case, 
the temperature gradient probed by the two CO lines is very small,
as both transitions are formed at nearly the same location in the disk.
In the case where D$_j$ is 10, however,
the $\tau=1$ surfaces for both transitions show an abrupt change 
near 130--150 AU in radius and 20--40 AU in height. 
With an even larger D$_j$ of 100, the CO J=3--2 line 
becomes optically thick significantly deeper (colder) in the disk than does
the CO J=2--1 line in the region beyond 200 AU where the temperature
is very low. The T$_B$ of CO 3--2 is predicted to be smaller than that 
of CO 2--1 in this case, which is not compatible with the SMA spectra 
reported here.

The observed difference in the brightness temperatures of the CO lines in 
Figure~\ref{fig:spectra} must therefore arise from a combination of factors.
The beam size for the CO J=3--2 spectrum corresponds to 
$\sim$160$\times$120 AU (FWHM), that for CO J=2--1 is
$\sim$260$\times$150 AU (FWHM). Both zones have a $\tau=1$ surface
above 22 K, though the 2--1 data do sample colder regions where
the $\tau=1$ surface drops to $<$20 K.
The velocity-integrated intensities for the full disk in the two CO lines
are $37.2\pm7.5$ Jy~km~s$^{-1}$ for CO 3--2 and 
$12.4\pm1.0$ Jy~km~s$^{-1}$ for CO 2--1,
as compared to previous single dish
observations of 36.6 Jy~km~s$^{-1}$ for CO 3--2 (\citealp{vanzadelhoff_v01}) 
and 17.7 Jy~km~s$^{-1}$ for CO 2--1 (\citealp{kastner_z97}). We
regard these as consistent given the uncertainties in the flux scales
of the various telescopes.  The ratio of integrated intensities from
the aggregate 345/230 GHz data sets is $2.45\pm0.6$, consistent with the
value of 2.25 expected if the lines were optically thick and tracing
isothermal gas. As mentioned above, even with the same convolving beam
the CO J=3--2 emission remains brighter than that from CO J=2--1 
(21 K vs 15 K). Figure~\ref{fig:spectra} shows that there is likely
to be a small vertical temperature gradient to which these two
transitions can be quite sensitive, e.g. for the
case of an inclination angle of 7$^\circ$ where CO J=2--1 
traces gas near $\sim$15 K, while the more optically thick
CO J=3--2 emission arises from gas closer to the disk surface
and therefore somewhat warmer ($\sim$18 K).
Given the likely flux uncertainties of the SMA and single
dish observations, it is difficult at present to quantify
whether vertical temperature gradients in the disk or beam
dilution dominate the observed difference in the brightness
temperatures of the CO lines shown in Figure~\ref{fig:spectra}.  
Studies of additional
CO transitions and, especially, isotopically substituted variants
could settle this issue definitively.

\begin{figure}
\centering
\includegraphics[width=2in,angle=-90]{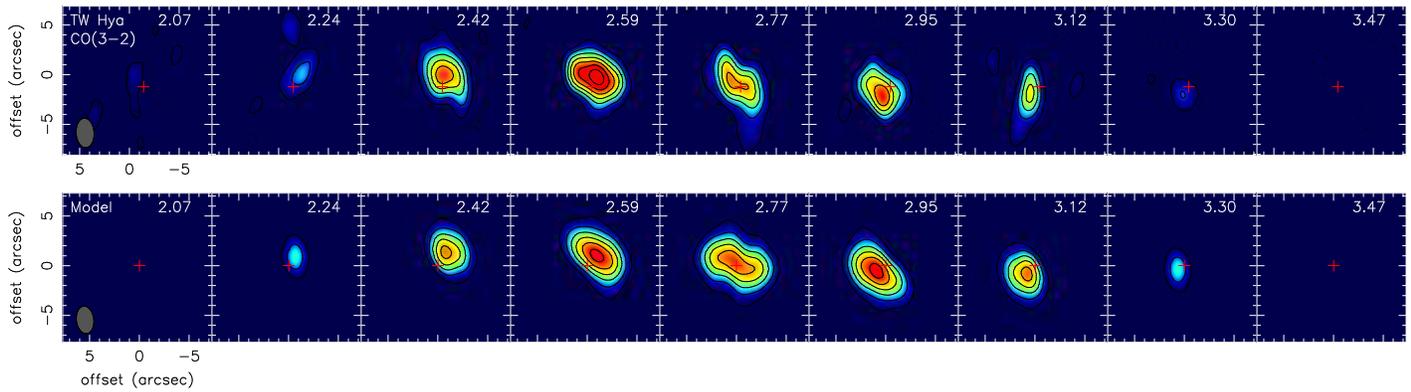}
\caption{
SMA images of the CO J=3--2 emission from TW Hya in
0.18 km~s$^{-1}$ velocity bins observed with the SMA
({\em upper panels}, with a contour spacing of 5 Jy~beam$^{-1}$), 
together with images from a simulated observation
using the model parameters listed in Table 2 ({\em lower panels},
with a contour spacing of 3 Jy~beam$^{-1}$). The small cross
indicates the position of the continuum source.
 \label{fig:cube}}
\end{figure}

Figure~\ref{fig:cube}~(upper panels) shows a set of velocity channel maps
for 0.18 km~s$^{-1}$ intervals for CO J=3--2 line. The main feature
of interest is the position shift along a northwest-southeast axis
at velocities that bracket the systemic velocity.
In order to interpret these images in a quantitative way, we use 
a two-dimensional accelerated Monte Carlo model (\citealp{hogerheijde_v00}) 
to calculate the radiative transfer and molecular excitation.
We adopt the physical density and temperature structure
derived by Calvet et al. (2002), which does an excellent job 
reproducing the dust continuum emission, and
produce a grid of models with a range of inclinations and 
various depletion scheme to simulate the disk as imaged by
a telescope with the resolution constrained only by the grid
sampling (typically this is of order 5-10 AU in the outer disk,
or $0\farcs1-0\farcs2$). Simulated observations of the model disks were
produced by the MIRIAD software package using the UVMODEL routine
to select synthetic visibility observations at the observed $(u,v)$ spacings.
These synthetic visibilities were then processed into images in 
a manner identical to the SMA data for detailed comparisons
(more detailed explanations and a description of the model may be found
in \citealp{kessler03}, for an application to LkCa 15 see Qi et al. 2003). 

The CO data provide a strong constraint on the disk inclination $i$, or
more specifically sin $i$. Previous analyses of scattered
light images (\citealp{krist_s00,trilling_k01}) indicated that TW Hya
and its disk are seen close to pole-on, but the inclination angle
could not be well constrained. For the disk models used here,
we fix the inner radius at 4 AU and the outer radius
at 196 AU, and adopt Keplerian rotation along with a low turbulent
velocity of 0.05 km~s$^{-1}$ through comparison to the observed
CO J=3--2 and J=2--1 line profiles. Indeed turbulent
velocities in the outer disk also affect the line width, but
values of $\geq$0.1 km~s$^{-1}$ or larger cannot reproduce the
peanut-shaped morphology of the systemic velocity channel map while the
line widths are essentially unchanged for turbulent velocities
in the 0.05--0.1 km~s$^{-1}$ range. Figure~\ref{fig:spectra} shows
the predictions for a range of inclination angles, which strongly
impact the widths of the lines; the best fit inclination is 7$^{\circ}$,
which is slightly larger than the inclination angle
of 4$^{\circ}$ derived from the ellipticity of the infrared image of the
disk around TW Hya in scattered light (\citealp{weinberger_b02}) where a
circular disk is assumed. Specifically, the predicted line widths change
by 30\%, from 0.72 to 1.06 km~s$^{-1}$ for CO J=3--2 and from 0.6-0.85
km~s$^{-1}$ for CO J=2--1, as the inclination angles vary from 6 to 8
degrees. Such significant line width variations with inclination
angle (by as little as 1 degree) are expected when the disk is so
close to pole-on because the locii of constant projected velocities is
quite sensitivite to sin $i$ -- the error of which translates to a
very tight limit on $i$.
An inclination angle of $7^{\circ}\pm 1^{\circ}$ is thus a robust
fit that is strongly constrained by the CO line widths within the
confines of the disk model employed. In general, the derived
inclination angle can also be coupled to the size of the disk
and the radial and vertical temperature gradients. The line intensity 
of CO 3--2 is
still much larger than the model prediction, suggesting that the
near-surface gas temperature in the Calvet model is somewhat too low.
The line profiles are also not
perfectly characterized as Gaussian (though they are very narrow),
which we rely on for determining the line widths.
Each these systematic uncertainties could be larger than
the uncertainty of the inclination angle (1 degree) that we
determine from the line widths, but can be constrained in the
future by higher spatial resolution observations of additional
CO transitions and isotopologues.

A comparison of the observed channel maps of CO J=3--2 and 
synthetic maps generated by a detailed model with parameters 
listed in Table~\ref{tab:model} is shown in Figure~\ref{fig:cube}. 
While the model reproduces all of the major 
features observed, in particular the velocity gradient and peanut-shaped 
morphology of the central channel, the predicted CO 3--2 intensities
in Figures~\ref{fig:spectra}~and ~\ref{fig:cube}~are 40\% smaller
than observed. It is likely that
vertical gas and dust temperature gradient difference is not quite 
captured by the SED-constrained disk models, which we will pursue
further in a future paper.
Resolved observations of CO isotopologues toward another T Tauri star,
DM Tau, have been presented by \citet{dartois_d03}, who used
IRAM inteferometric data to constrain the depletion and temperature
versus radius, as the less abundant species probe the inner part
of the disk, within an LTE excitation/radiative
transfer model. Such an approach is useful for molecules such as
CO that possess low critical densities, but as has been discussed
by \citet{vanzadelhoff_a03}, LTE excitation is not a
good approximation for higher dipole moment species such as CN.
With the 2D Monte Carlo model, the statistical nature of the
molecular excitation can be investigated within the context of
realistic models of the disk structure. Chemical differentiation
can therefore be constrained by resolved observations
of multiple transitions of species with varying characteristics,
and our future publications will thus include studies of other
molecules, including CN and HCN, toward TW Hya.

We thank all the SMA staff members for their diligent work in 
completing the SMA. We also thank P. D'Alessio and N. Calvet 
for providing the disk model of TW Hya and J.E. Kessler-Silacci for 
helpful discussions concerning the radiative transfer model. And we 
are grateful to J. H. Kastner for useful comments.

\bibliographystyle{apj}


\begin{deluxetable}{lcc}
\tablewidth{0pt}
\tablecaption{Observational Parameters for SMA TW Hya \label{tab:obs}}
\tablehead{
\colhead{}&\colhead{CO 3--2}&\colhead{CO 2--1}}
\startdata
Rest Frequency:& 345.796 GHz & 230.538 GHz \\
Observations:  & 2003 Mar 15 & 2003 Mar 24 \\ 
               & 2003 Apr 21 & 2003 Apr 20 \\
               &             & 2003 May 03  \\
Synthesized beam: & $2\farcs9 \times 2\farcs1$ PA 3.4$^{\circ}$ 
                  & $4\farcs7 \times 2\farcs7$ PA 9.9$^{\circ}$ \\
R.M.S$\tablenotemark{a}$ (continuum): &   35 mJy/beam  &  6.8 mJy/beam \\
Dust flux:    & 1.46 $\pm$ 0.04Jy & 0.57 $\pm$ 0.02Jy \\
Channel spacing: & 0.18 km s$^{-1}$ & 0.26 km s$^{-1}$ \\
R.M.S.$\tablenotemark{a}$ (line):   & 1.1 Jy/beam & 0.28 Jy/beam \\
Integrated intensity & 23.7 K km s$^{-1}$ & 10.6 K km s$^{-1}$ \\
Peak intensity & 25.9 K & 14.7 K \\
$\Delta V$\tablenotemark{b} & $0.86\pm0.02$ km s$^{-1}$ & $0.73\pm0.02$ km s$^{-1}$ \\
\enddata 
\tablenotetext{a} { SNR limited by the dynamic range. }
\tablenotetext{b} { Based on Gaussian fits to the spectra (FWHM). }               
\end{deluxetable}

\begin{deluxetable}{lcccc}
\tablewidth{0pt}
\tablecaption{Model Parameters Used in Simulating TW Hya CO emission \label{tab:model}}
\tablehead{
\colhead{} &\colhead{Parameters}}
\startdata
Physical Structure & Irradiated accretion disk (Calvet et al. 2002)\\
Stellar Mass & 0.6 M$_{\odot}$\\
Disk Size & R$_{in}$ 4 AU, R$_{out,edge}$ 196 AU\\
Disk PA   & 45$^{\circ}$ \\
Inclination Angle & 7$^{\circ}$ \\
Turbulent Velocity & 0.05 km s$^{-1}$\\
Depletion Factor & 10$\times$ (100$\times$ for T$\le$22 K)\\
\enddata
\end{deluxetable}


\newpage
\clearpage

\clearpage

\end{document}